\RequirePackage{iftex}\ifPDFTeX\pdfoutput=1\fi
\documentclass[11pt,a4paper]{article}

\usepackage[T1]{fontenc}
\usepackage[utf8]{inputenc}
\IfFileExists{lmodern.sty}{\usepackage{lmodern}}{\usepackage{ae,aecompl}}

\usepackage{amsmath,amssymb,amsthm}
\usepackage{mathtools}

\usepackage[margin=2.7cm]{geometry}
\usepackage[expansion=false]{microtype}
\usepackage{booktabs}
\usepackage{enumitem}

\usepackage{xcolor}

\usepackage{listings}
\definecolor{leankeyword}{RGB}{0,64,128}
\definecolor{leancomment}{RGB}{96,96,96}
\definecolor{leanstring}{RGB}{128,0,64}

\lstdefinelanguage{lean4}{
  keywords={theorem,lemma,def,structure,inductive,where,by,decide,
            import,open,namespace,end,variable,fun,match,with,
            deriving,instance,if,then,else,let,have,show,sorry},
  sensitive=true,
  comment=[l]{--},
  morecomment=[s]{/-}{-/},
  string=[b]",
}

\usepackage[colorlinks=true,linkcolor=blue!50!black,citecolor=blue!50!black,urlcolor=blue!50!black]{hyperref}
\usepackage[capitalise,noabbrev]{cleveref}

\newcommand{\Real}{\mathbb{R}}
\newcommand{\Rat}{\mathbb{Q}}

\newcommand{\Rthree}{\Real^3}
\DeclareMathOperator{\dotp}{dot}
\DeclareMathOperator{\crossp}{cross}
\DeclareMathOperator{\orthG}{orthGraph}
\newcommand{\vecv}[1]{\mathbf{v}_{#1}}
\newcommand{\basis}[1]{\mathbf{e}_{#1}}
\newcommand{\normsq}[1]{\lVert #1 \rVert^2}
\newcommand{\propext}{\textnormal{\lstinline|propext|}}
\newcommand{\Choice}{\textnormal{\lstinline|Classical.choice|}}
\newcommand{\QuotSound}{\textnormal{\lstinline|Quot.sound|}}
\newcommand{\axiomClosure}{\{\propext,\ \Choice,\ \QuotSound\}}
\newcommand{\checksound}{\lstinline|Cert.check_sound|}
\newcommand{\certcheck}{\lstinline|Cert.check|}
\newcommand{\physcheck}{PhysicsCheck}
\newcommand{\krat}{\lstinline|KRat|}
\newcommand{\uw}{Uijlen--Westerbaan}
\newcommand{\lbg}{Li--Bright--Ganesh}
\newcommand{\Lone}{\textsf{L1}}
\newcommand{\Ltwo}{\textsf{L2}}
\newcommand{\Lthree}{\textsf{L3}}
\newcommand{\Lfour}{\textsf{L4}}
\newcommand{\numSourceLines}{291}
\newcommand{\numDistinct}{180}
\newcommand{\numDistinctTwelve}{173}
\newcommand{\numAdversarial}{115}
\newcommand{\numSurvivors}{44}
\newcommand{\code}[1]{\texttt{#1}}

\newtheorem{theorem}{Theorem}[section]

\theoremstyle{definition}
\newtheorem{definition}[theorem]{Definition}
\newtheorem{finding}{Finding}
\theoremstyle{remark}

\title{Machine-Checked Certificates for the Geometric Half\\
       of the Minimum Kochen--Specker Bound}

\author{Shayaan Siddique \qquad Ibrahim Mian\\[2pt]
  \normalsize Millennium Research\\
  \normalsize\texttt{\{shayaan,ibby\}@millenniumresearch.ai}\\
  \normalsize\texttt{shayaansiddique02@gmail.com}, \texttt{ibrahimnmian@gmail.com}}

\date{}

\begin{document}
\maketitle

\begin{abstract}
The best known lower bound for the minimum Kochen--Specker vector system
in $\Rthree$---24 vectors---rests on a computational proof whose
combinatorial half emits DRAT proofs but whose geometric half does not:
the non-embeddability of thousands of candidate graphs is established by
Z3's nonlinear real arithmetic, which produces no checkable proof
objects.
We close this gap for the proof's blocking database.
We introduce exact rational \emph{case-tree certificates} of real
non-embeddability, whose splits are polynomial factorizations and
rational sum-of-squares decompositions and whose leaves are discharged by
injectivity, ideal-membership, or Positivstellensatz-shaped positivity
arguments, and we certify all $\numSourceLines$ source lines
($\numDistinct$ distinct graphs) of the published pipeline's
order-$10$--$13$ blocking lists.
Certificates are replayed by two independent checkers that share no code
with the generator: a pure-Python replay over exact fractions, and a
total checker implemented and \emph{proved sound} in Lean~4.
The soundness theorem---acceptance implies that no injective-on-rays,
orthogonality-respecting assignment of nonzero real vectors realizes the
graph---is kernel-checked with axiom closure $\axiomClosure$, and a
gcd-free rational arithmetic layer makes the entire verdict computation
kernel-reducible, so each per-graph non-embeddability result is a closed
kernel theorem proved by \lstinline|decide|.
The formalization surfaced findings about the published pipeline,
including a load-bearing injectivity side condition in its embeddability
notion, hidden WLOG case obligations invisible to Z3-based workflows, and
an unreproducible candidate count that we resolve against the published
artifacts.
All certificates, checkers, and proofs are available and replayable from
a single build.
\end{abstract}

% =========================================================================
\section{Introduction}\label{sec:intro}

A Kochen--Specker (KS) vector system in $\Rthree$ is a finite set of nonzero
vectors, pairwise distinct as rays, that admits no $\{0,1\}$-coloring in which
every orthogonal pair contains at most one $1$ and every orthogonal triad
contains exactly one $1$. Such systems witness the Kochen--Specker theorem
\cite{KochenSpecker1967}, a cornerstone of quantum foundations ruling out
noncontextual hidden-variable models, and the size of the smallest such system
is a long-standing open problem: the best construction has $31$ vectors, and
the best lower bound, established computationally by Li et al.\
\cite{LiBrightGanesh2024} and Kirchweger et al.\ \cite{KirchwegerPeitlSzeider2023},
is $24$.

The published lower-bound proof has two halves. The combinatorial half is an
isomorph-free SAT enumeration of candidate graphs up to order $23$; its
unsatisfiability results emit DRAT proofs checkable by \code{drat-trim}
\cite{WetzlerHeuleHunt2014}. The geometric half must show that every
enumerated candidate graph is non-embeddable in $\Rthree$---that no assignment
of nonzero, pairwise-ray-distinct vectors realizes its edges as
orthogonalities. This half is established by Z3's nonlinear real arithmetic
solver \cite{deMouraBjorner2008}, which produces no proof objects: the trust
chain for the geometric half of the bound terminates in an unverified decision
procedure for a $\exists\Real$-hard theory. The asymmetry is stark. One half
of a flagship computational result in quantum foundations is independently
checkable; the other must be taken on faith.

\paragraph{Contributions.}
We close this gap for the blocking database on which the published pipeline
leans, and we lay verified groundwork for the full bound. Specifically:

\begin{enumerate}[leftmargin=2em]
\item \emph{Certificate format (\cref{sec:certificates}).} We introduce exact
rational case-tree certificates of real non-embeddability. A certificate
WLOG-normalizes an anchor triad or edge, replays a propagation schedule that
expresses all vertex vectors in few parameters, and presents a case tree whose
splits are polynomial factorizations and rational sum-of-squares (SOS)
decompositions, with leaves discharged by injectivity collisions,
nonzero-obligation collapse, ideal membership with explicit cofactors, or a
Positivstellensatz-shaped positivity argument. Every inference is checkable by
exact $\Rat$-arithmetic; no numerical tolerance appears anywhere in the trust
path.
\item \emph{Complete coverage of the blocking database (\cref{sec:results}).}
We certify all $\numSourceLines$ source lines ($\numDistinct$ distinct graphs)
of \physcheck's order-10/11/12 minimal-non-embeddable blocking lists together
with the seven order-13 \uw{} minimal graphs needed for the final candidates,
replacing every Z3 verdict the database rests on.
\item \emph{Dual independent replay and a kernel-checked soundness theorem
(\cref{sec:lean}).} Certificates are validated by two checkers that share no
code with the generator: a pure-Python replay over exact fraction arithmetic,
and a total checker \certcheck{} implemented in the Lean~4 proof assistant
\cite{deMouraUllrich2021}. We prove \checksound: acceptance implies that no
injective-on-rays, orthogonality-respecting assignment of nonzero $\Rthree$
vectors realizes the certified graph. The development is \code{sorry}-free,
with an axiom closure of $\axiomClosure$ (no native reduction, no custom
axioms) enforced by a build-time axiom gate. A gcd-free rational layer
(\krat) and fuel-structural recursion make the verdict computation
kernel-reducible, so each per-graph non-embeddability corollary is a closed
kernel theorem proved by \lstinline|decide| in roughly half a second.
\item \emph{Findings about the published pipeline (\cref{sec:findings}).}
Formalization forced precision the original workflow did not: we show that
pairwise ray-distinctness is load-bearing in the published embeddability
notion (several database graphs are non-embeddable only for injectivity
reasons); that the orthoplane parametrization hides WLOG case obligations
invisible to Z3-based pipelines; that a noncolorability optimization treated
as a lemma is empirical and not load-bearing; and that the published count of
final order-23 candidates (41) is not reproducible from the published
artifacts under the documented filter---the correct count is 43, which we
resolve exactly against pinned artifact hashes.
\item \emph{Adversarial validation (\cref{sec:results}).} $\numAdversarial$
certificate mutations---dropped branches, tampered coefficients, incomplete
case splits, malformed monomials---are rejected by both checkers in continuous
integration, on the same code paths that accept the genuine certificates.
\end{enumerate}

Everything in this paper replays from a single build: the Python suite
requires only a Python $\ge 3.9$ interpreter, and the Lean development builds
with a pinned toolchain and pinned Mathlib against \code{\#guard}-checked
certificate transcriptions.

\paragraph{Scope.}
This paper verifies the geometric half for the \emph{blocking database}: the
graphs whose non-embeddability the published enumeration consults at every
order, plus the order-13 graphs that discharge $42$ of the $44$ final
candidates by verified subgraph containment. It does not yet compose a
verified end-to-end bound $\ge 24$ theorem; \cref{sec:fullbound} lays out the
remaining layers---encoding correctness, verified canonicity, and streaming
LRAT replay at terabyte scale---and the design decisions each one forces.

\begin{table}[t]
\centering\small
\begin{tabular}{l l l l}
\toprule
 & Component & Basis in published proof & Status here \\
\midrule
T1 & CNF encoding correctness & hand proof in paper & analyzed (\cref{sec:findings}) \\
T2 & minimality WLOGs & hand proofs & lemma inventory (\cref{sec:lean}) \\
T3 & canonicity pruning & trusted solver code & route fixed (\cref{sec:fullbound}) \\
T4 & SAT unsatisfiability & DRAT via \code{drat-trim} & streaming plan (\cref{sec:fullbound}) \\
T5 & non-embeddability & Z3 NRA, no proof objects & \textbf{closed for blocking DB} \\
\bottomrule
\end{tabular}
\caption{Trust surface of the published lower-bound pipeline. This paper
closes T5 for the blocking database (bold); \cref{sec:fullbound} addresses the
remainder.}
\label{tab:trust}
\end{table}

% =========================================================================
\section{Background and Trust Surface}\label{sec:background}

\subsection{The minimum Kochen--Specker problem}

Fix nonzero vectors $v_1, \dots, v_n \in \Rthree$, pairwise distinct as rays.
Their \emph{orthogonality graph} $\orthG$ has a vertex per vector and an edge
per orthogonal pair. A \emph{010-coloring} assigns $\{0,1\}$ to vertices so
that adjacent vertices are not both $1$ and every triangle receives exactly
one $1$; a \emph{KS system} is a vector system whose orthogonality graph
admits no 010-coloring. The \emph{minimum KS problem} asks for the least $n$
admitting such a system. Uijlen and Westerbaan \cite{UijlenWesterbaan2016}
established $n \ge 22$; Li et al.\ \cite{LiBrightGanesh2024} and Kirchweger et
al.\ \cite{KirchwegerPeitlSzeider2023} independently raised the bound to
$n \ge 24$ using SAT-driven enumeration, and $31$ remains the best upper bound
\cite{ConwayKochen1993}.

The enumeration strategy is common to both proofs: generate, up to
isomorphism, all KS candidate graphs of order $\le 23$---graphs that are
squarefree, 010-noncolorable after symmetry and minimality pruning---and show
that none embeds in $\Rthree$ as an orthogonality graph of distinct rays. If
no candidate embeds, no KS system of that order exists.

\subsection{Trust surface of the published proof}

\cref{tab:trust} summarizes the trust basis of each component of the \lbg{}
pipeline (\physcheck), from our source audit. The SAT half (T4) is the
strongest link: unsatisfiability results carry DRAT proofs, albeit checked by
an unverified C checker. The encoding (T1), minimality arguments (T2), and
canonicity pruning (T3) rest on hand proofs and trusted solver internals. The
headline gap is T5: non-embeddability of candidates is decided by Z3's
nonlinear real arithmetic, plus a Python forced-assignment routine and
blocking lists of minimal non-embeddable subgraphs---themselves Z3-derived. No
component of T5 emits a proof object.

\subsection{Why proof objects are hard here}

The non-embeddability statements live in the existential theory of the reals:
a graph embeds iff a system of quadratic orthogonality equations,
non-vanishing constraints, and ray-distinctness disequalities has a real
solution. Z3 decides such systems, but its NRA engine is a complex decision
procedure whose runs are not reconstructible as checkable objects. Worse,
several obstructions in this database are \emph{real-geometric}: the
orthogonality variety is nonempty over $\mathbb{C}$, and non-embeddability
holds only because every real point collapses a pair of rays. Complex ideal
membership (a Gr\"obner-basis certificate alone) is provably insufficient for
these graphs; any complete certificate format must speak about real
positivity. Our format does so through rational SOS splits and a
Positivstellensatz-shaped positivity leaf (\cref{sec:certificates}), while
keeping every check inside exact $\Rat$-arithmetic---the checkers never
factor, never search, and never call a CAS.

% =========================================================================
\section{Case-Tree Certificates of Real Non-Embeddability}\label{sec:certificates}

A certificate for a graph $G$ is a self-contained object---anchor, propagation
schedule, parameter list, and case tree---replayed by a checker that performs
only exact rational arithmetic on structured polynomials (monomial lists over
$\Rat$, mirroring the Lean \lstinline|MPoly| normal form). The generator may
be clever; the checkers stay dumb, total, and exact.

\subsection{Embedding notion}

Formalization begins by fixing what ``embeddable'' means, and here the
published pipeline holds a surprise (Finding~\ref{find:injectivity}):
\physcheck's Z3 encoding asserts pairwise ray-distinctness for \emph{all}
vertex pairs, not merely adjacent ones, making injectivity part of
embeddability. Our verified statement therefore builds it in.

\begin{definition}[Realization]\label{def:realization}
A \emph{realization} of a graph $G$ on vertices $1, \dots, n$ is a map
$v \mapsto \vecv{v} \in \Rthree$ such that (i) $\vecv{v} \ne 0$ for all $v$;
(ii) $\dotp(\vecv{i}, \vecv{j}) = 0$ for every edge $\{i,j\}$; and (iii) for
all $v \ne w$, $\vecv{v}$ and $\vecv{w}$ are distinct as rays. $G$ is
\emph{non-embeddable} if it has no realization.
\end{definition}

\subsection{WLOG normalization and the propagation schedule}

The rotation group acts transitively on ordered orthonormal triads, and on
ordered orthonormal pairs; per-vertex scaling is free. A certificate opens by
anchoring either an orthogonal triangle $(a,b,c) \mapsto
(\basis{1},\basis{2},\basis{3})$ or, for triangle-free graphs, an edge
$(a,b) \mapsto (\basis{1},\basis{2})$. It then replays a schedule of two step
forms:

\begin{itemize}
\item \code{cross(v, u, w)}: assign $\vecv{v} \coloneqq \vecv{u} \times
\vecv{w}$ where $v$ is adjacent to both assigned vertices $u \ne w$. Soundness
is the orthocomplement lemma ($w \perp u$, $w \perp v$, $u \nparallel v$ imply
$w \parallel u \times v$), and the side condition $\vecv{u} \times \vecv{w}
\ne 0$ becomes a recorded \emph{obligation polynomial}
$\normsq{\vecv{u} \times \vecv{w}}$, justified by ray-distinctness of $u$ and
$w$.
\item \code{plane(v, u, axis)}: parametrize $\vecv{v} \coloneqq s_v\,p +
t_v\,q$ with $p = \vecv{u} \times \basis{\mathrm{axis}}$ and
$q = \vecv{u} \times p$, introducing fresh parameters $s_v, t_v$ and
obligations $\normsq{p}$ and $s_v^2 + t_v^2$.
\end{itemize}

The orthoplane parametrization conceals a case the informal argument never
surfaces: the spanning pair $(p, q)$ degenerates when the parent vector is
parallel to the chosen basis vector. That branch must be discharged by
injectivity against an anchor vertex, which is why edge-anchored certificates
restrict \code{plane} axes to anchor-owned axes ($\basis{1}, \basis{2}$, never
$\basis{3}$)---the sound discharge is otherwise unavailable
(Finding~\ref{find:wlog}).

After the schedule, every vertex vector is a polynomial in the parameters; the
\emph{pool} consists of one orthogonality polynomial $\dotp(\vecv{i},
\vecv{j})$ per induced edge, shrunk by inline linear elimination steps whose
provenance the checker re-derives (each substitution names the pool polynomial
it comes from), typically leaving at most two residual polynomials.

\subsection{The case tree}

Internal nodes split the current branch; leaves close it. Splits take two
forms: a \emph{factorization split} on $p = \prod_k f_k$ (in an integral
domain, a vanishing product forces some factor to vanish---the checker
verifies the product identity by expansion), and an \emph{SOS split} on
$p = \sum_k c_k m_k^2$ with $c_k \in \Rat_{>0}$ (over a formally real field, a
vanishing sum of squares forces every summand to vanish). SOS witnesses are
found offline by exact Gram-matrix $LDL^T$ decomposition over an explicit
monomial basis, but the certificate records only $(c_k, m_k)$ and the checker
verifies the identity by expansion; the discovery procedure is not trusted.
Leaves are:

\begin{itemize}
\item \code{collide(i, j)}: the replayed vectors of distinct vertices $i, j$
are provably parallel on this branch, contradicting ray-distinctness;
\item \code{oblig}: an obligation polynomial (or a designated basis component
of an anchor-adjacent vector) is identically zero on this branch,
contradicting its justification;
\item \code{ideal} and \code{one}: explicit cofactors $q_1, \dots, q_m$ with
$\sum_i q_i p_i = g$ over the residual pool $\{p_1, \dots, p_m\}$, where $g$
is an obligation polynomial or a pair-injectivity polynomial
$\normsq{\vecv{i} \times \vecv{j}}$ (\code{ideal}), or $\sum_i q_i p_i = 1$
(\code{one}: the pool has no common zero over any field);
\item \code{positiv}: cofactors $q_r$, vertex pairs
$(i_1, j_1), \dots, (i_m, j_m)$, and an SOS list $(c_k, h_k)$ with the
polynomial identity
\begin{equation}\label{eq:positiv}
\prod_{l=1}^{m} \normsq{\vecv{i_l} \times \vecv{j_l}}
+ \sum_k c_k h_k^2
= \sum_r q_r \cdot p_r,
\end{equation}
verified by expansion. At a real point of the branch the right side vanishes
while the left is strictly positive---each pair factor is positive by
injectivity and the SOS term is nonnegative---a contradiction.
\end{itemize}

The \code{positiv} leaf is the Positivstellensatz-shaped generalization of
\code{ideal}, and it is not a luxury: some branch varieties are real-nonempty,
but every real point collapses a \emph{varying} vertex pair, so no single
collision or ideal argument closes the branch.

Finally, a certificate may carry a \code{restrict\_to} field naming an induced
subgraph $H \subseteq G$ actually certified; subgraph monotonicity (a
realization of $G$ restricts to one of any subgraph, induced or not) lifts
non-embeddability to $G$. This is needed because the published order-11/12
blocking lists contain graphs consisting of a non-embeddable core plus
isolated vertices, and---in its general, non-induced form---it is exactly the
lemma that lets seven order-13 certificates discharge $42$ of the
$\numSurvivors$ final candidates (\cref{sec:results}).

\subsection{A complete example}

\cref{fig:example} shows a certificate for the first order-10 graph in the
database, verbatim from the human-readable rendering. The tree is three levels
deep and closes every branch by injectivity.

\begin{figure}[t]
\begin{lstlisting}[language={}]
SPLIT[FACTOR] on -s2^6 s6 t6^2 - s2^4 s6^3 - 2 s2^4 s6 t2^2 t6^2
  [s6 = 0]  LEAF COLLIDE (1,3)   rays of vertices 1,3 coincide
  [s2 = 0]  LEAF OBLIG basis[2,e1]   vertex 2 parallel to anchor
  [s2^2 t6^2 + s6^2 + 2 t2^2 t6^2 = 0]
            SOS => all summands 0 => LEAF COLLIDE (1,3)
\end{lstlisting}
\caption{Certificate for order-10 graph \#1 (verbatim). The factorization
split is verified by expanding the product; the SOS split by expanding
$\sum_k c_k m_k^2$; both leaves are injectivity contradictions.}
\label{fig:example}
\end{figure}

\subsection{Schema and normative deserialization}

Certificates ship as JSON (\code{schema\_version = 1}) with fully structured
polynomials: a monomial is a rational coefficient with positive denominator
and a duplicate-free, exponent-positive variable list. Well-formedness is
normative and enforced at ingestion by both checkers; the adversarial suite
includes a duplicated-variable monomial that normalizes to the correct
polynomial, so only syntactic rejection distinguishes it
(\cref{sec:results}). The replay checkers share no code and no parsing with
the generator.

% =========================================================================
\section{A Kernel-Checked Soundness Theorem in Lean 4}\label{sec:lean}

The Lean development (toolchain v4.32.0, pinned Mathlib) implements the
checker as a total reflective function \certcheck{} over the same normal-form
polynomial arithmetic the schema prescribes, and proves it sound against the
semantic embedding notion of \cref{def:realization}.

\subsection{The soundness theorem}

\cref{fig:soundness} gives the statement. The proof factors through an
\emph{interpretation invariant} for the schedule replay (orthocomplement
representatives for \code{cross} steps; well-posed $2 \times 2$ solves for
\code{plane} steps), an elimination replay that consumes exactly what the
checker verifies syntactically, and a descent through all eight node kinds of
the case tree. The real-geometry layer supplies the seven lemmas the tree walk
consumes: transitivity of the $O(3)$ action on anchored triads and edges,
per-vertex scaling, the orthocomplement lemma, injectivity of the cross
product on distinct rays, plane-basis spanning with its parallel branch,
SOS-zero over formally real fields, and factor-zero in integral domains; the
leaf extensions add evaluation of ideal identities (evaluation is a ring
homomorphism) and the positivity argument for \cref{eq:positiv}. The full
lemma inventory, including the combinatorial-layer lemmas (squarefreeness of
orthogonality graphs, noncolorability of KS graphs, vertex deletion, subgraph
monotonicity), is maintained in the repository with the invariant that any
change to certificate semantics lands only after its justifying lemma is
inventoried.

\begin{figure}[t]
\begin{lstlisting}[language=lean4]
/-- A real realization of the graph carried by certificate `c`:
    nonzero vectors for the vertices `1..c.n`, orthogonal along every
    edge, pairwise distinct as rays. -/
structure Realization (c : Cert) where
  vec      : Nat → Vec3 ℝ
  nonzero  : ∀ v, 1 ≤ v → v ≤ c.n → vec v ≠ ⟨0, 0, 0⟩
  orth     : ∀ e ∈ c.edges, dot (vec e.1) (vec e.2) = 0
  distinct : ∀ v w, 1 ≤ v → v ≤ c.n → 1 ≤ w → w ≤ c.n → v ≠ w →
               ¬ SameRay (vec v) (vec w)

/-- Soundness: checker acceptance refutes every realization. -/
theorem check_sound (c : Cert) (h : check c = true)
    (x : Realization c) : False
\end{lstlisting}
\caption{The verified statement. \lstinline|check_sound| is
\lstinline|sorry|-free with axiom closure $\axiomClosure$.}
\label{fig:soundness}
\end{figure}

Discipline is enforced mechanically. A build-time axiom gate audits the axiom
closure of every declaration (no axioms beyond $\axiomClosure$, no native
reduction), synchronizes the \code{sorry} set against a manifest in both
directions---the manifest is empty---and confines Mathlib imports to the
real-geometry subtree; the combinatorial and checker layers build against core
and Batteries only.

\subsection{Kernel-reducible verdicts}

A soundness theorem transfers no trust to a graph until the hypothesis
\lstinline|check c = true| is itself established inside the trusted kernel.
Initially this was only possible by compiled evaluation (\code{\#guard}), for
two structural reasons: core \lstinline|Rat| normalizes through
\lstinline|Nat.gcd|, which does not reduce feasibly in the kernel, and the
checker's well-founded recursions do not unfold definitionally.

We closed the gap without \lstinline|native_decide| and without changing
semantics. The \krat{} layer provides gcd-free, normalization-free fractions:
integer numerator/denominator pairs carrying a $\mathrm{den} \ne 0$ invariant
maintained by the operations, with zero-testing by numerator, equality by
cross-multiplication, and sign via $\mathrm{num} \cdot \mathrm{den}$. The
polynomial merge and the tree walk were rewritten in fuel-structural form with
evaluation-preserving (append) and rejecting (\lstinline|false|) fallbacks
respectively, so every semantics lemma holds at every fuel value and fuel
exhaustion can only refuse a certificate---never accept one. The whole verdict
computation is then kernel-reducible, and each per-graph corollary is a closed
kernel theorem:

\begin{lstlisting}[language=lean4]
theorem c10_0_nonembeddable : /- no Realization of graph #0 -/ := by decide
\end{lstlisting}

Kernel evaluation costs $\approx 0.5$\,s per order-$\le 12$ verdict and
$\approx 1.8$\,s per order-13 verdict on a laptop (\cref{tab:timings}); the
escalation threshold budgeted in the phase plan ($60$\,s per certificate) was
never approached. Each generated theorem pins its graph to an explicit colex
edge list by \code{\#guard}, independently re-verified against the \physcheck{}
source files by the Python replay, so the binding between a source line and
its theorem is itself checked, not asserted. Certificate transcription into
Lean is deterministic and diff-checked in CI, and the coverage table in the
repository README is regenerated from the compiled Lean environment, not from
the JSON.

% =========================================================================
\section{Results}\label{sec:results}

\subsection{Coverage}

\begin{table}[t]
\centering\small
\begin{tabular}{l r r r c r}
\toprule
order & source lines & distinct & certified & replay & kernel thms \\
\midrule
10 & 2 & 2 & 2/2 & \checkmark & 2/2 \\
11 & 16 & 16 & 16/16 & \checkmark & 16/16 \\
12 & 266 & 155 & 266/266 & \checkmark & 155/155 \\
13 (\uw{} minimal) & 7 & 7 & 7/7 & \checkmark & 7/7 \\
\midrule
total & \numSourceLines & \numDistinct & \numSourceLines/\numSourceLines & \checkmark & \numDistinct/\numDistinct \\
\bottomrule
\end{tabular}
\caption{Certification coverage of the \physcheck{} blocking database and the
order-13 \uw{} minimal graphs. ``Source lines'' counts lines of the published
graph lists; duplicate lines share their graph's theorem, with the line--graph
binding verified by replay.}
\label{tab:coverage}
\end{table}

\cref{tab:coverage} summarizes coverage: every distinct graph in the blocking
database carries a closed kernel non-embeddability theorem, and every source
line is bound to its theorem. The order-10 trees are at most three levels deep
and fully human-readable; deeper trees at orders 11--13 exercise edge anchors,
induced-subgraph restriction, and the ideal and positivity leaves. The seven
order-13 certificates were produced in 18 seconds of generator time, a
consequence of the format's maturity by that phase.

The order-13 row does real work for the full bound. Replaying the published
pipeline's candidate artifacts (Finding~\ref{find:count}), the final
order-22/23 candidate set has $\numSurvivors$ members after the pipeline's own
$\le 12$-subgraph filter---and none of them contains any of our
$\numDistinctTwelve$ certified order-$\le 12$ graphs, so containment against
the existing database discharges nothing. But exactly $42$ of the
$\numSurvivors$ contain one of seven order-13 \uw{} minimal non-embeddable
graphs, so certifying those seven discharges $42$ candidates via the verified
general (non-induced) subgraph-monotonicity lemma. The remaining two
candidates are recorded machine-readably as unresolved obligations---never
faked, appearing as explicit hypotheses in any composed statement until
closed---with direct certification in progress.

\subsection{Kernel timings}

\begin{table}[t]
\centering\small
\begin{tabular}{l r r r}
\toprule
shard & certificates & marginal user time & per certificate \\
\midrule
order-10 shard & 2 & 0.9\,s & $\approx 0.45$\,s \\
order-12 shard & 20 & 10.0\,s & $\approx 0.50$\,s \\
order-13 shard (\uw) & 7 & 12.4\,s & $\approx 1.8$\,s \\
\bottomrule
\end{tabular}
\caption{Marginal kernel cost of \lstinline|decide| verdicts (laptop-class
arm64, Lean 4.32.0). Marginal user CPU time over an imports-only baseline;
wall time is dominated by $\approx 76$\,s of olean loading.}
\label{tab:timings}
\end{table}

\cref{tab:timings} reports kernel-evaluation costs. Extrapolated total kernel
and elaboration cost for all $\numDistinctTwelve$ order-$\le 12$ certificates
is roughly $90$\,s of CPU; the full ten-shard build completes in minutes under
parallel contention. Kernel-checking the geometric half of this database is,
in short, cheap enough to run in CI on every commit---which it does.

\subsection{Adversarial validation}

Checker code that has only ever seen valid certificates is untested where it
matters. The adversarial suite mutates genuine certificates in
$\numAdversarial$ ways---dropping a branch of a split, tampering with an SOS
coefficient, truncating a factorization, removing a leaf justification,
injecting a duplicated-variable monomial that normalizes correctly, corrupting
cofactors of an ideal identity, breaking the edge-list binding---and requires
both checkers to reject every mutant. The Lean side runs the mutations through
the same \certcheck{} code path used by the kernel theorems (via a replay
executable), giving CI-demonstrated adversarial parity between the two
checkers: $\numAdversarial$/$\numAdversarial$ rejected by both.

% =========================================================================
\section{What Formalization Revealed About the Published Pipeline}\label{sec:findings}

A recurring theme of verified re-examinations of computational proofs is that
the formalization pays for itself in findings. Ours is no exception.
Throughout, the governing rule was \emph{soundness over coverage}: whenever a
proof needed a hypothesis the checker did not verify syntactically, the
resolution was to strengthen the checker or report a finding---never to weaken
a check or paper over a discrepancy.

\begin{finding}[Injectivity is load-bearing]\label{find:injectivity}
\physcheck's Z3 encoding asserts $\crossp(v_j, v_i) \ne 0$ for \emph{all}
pairs $i \ne j$, making pairwise ray-distinctness part of its embeddability
notion. This is not cosmetic: several database graphs have orthogonality
systems with real solutions, every one of which collapses some (varying)
vertex pair. Those graphs are non-embeddable \emph{only} for injectivity
reasons. Any verified statement must therefore build injectivity into the
embedding definition---ours does (\lstinline|Realization.distinct|)---and any
complete certificate format must reason about injectivity loci, not just the
orthogonality ideal, which is what forced the \code{positiv} leaf of
\cref{sec:certificates}.
\end{finding}

\begin{finding}[Hidden WLOG case obligations]\label{find:wlog}
The orthoplane parametrization conceals a degenerate branch---parent vector
parallel to the chosen basis vector---that must be discharged by injectivity
against an anchor vertex, and the edge-anchor variant is sound only because
\code{plane} axes are restricted to anchor-owned axes. Z3-based pipelines
never surface such case obligations; a checker with a soundness theorem cannot
avoid them.
\end{finding}

\begin{finding}[The half-ones optimization is empirical, not a
lemma]\label{find:halfones}
The pipeline's noncolorability encoding blocks only colorings with
$1 \le |\text{ones}| \le \lceil n/2 \rceil$. Our verification plan initially
treated the underlying claim as a load-bearing unverified lemma; the audit
overturned this premise. The paper itself justifies the restriction
empirically, and it is not load-bearing for the bound: dropping blocking
clauses only enlarges the candidate set, and the direction the verified chain
needs---a noncolorable graph satisfies every blocking clause---is trivial and
proved \code{sorry}-free. The stronger counting statement remains open (the
natural injection proof provably fails on a girth-5 edge-count obstruction)
and is recorded as a proof-free conjecture, relevant only to completeness
claims, not soundness. Two secondary paper-versus-code discrepancies are
documented for the encoding-correctness layer: an off-by-one in the blocking
range (paper text $< \lceil n/2 \rceil$, code $\le \lceil n/2 \rceil$) and the
never-blocked empty coloring (benign only because the triangle constraint
forces a triangle). Relatedly, one documented minimality constraint is
commented out in the generator source, so the encoding-correctness target must
match the constraint set \emph{as shipped}, not as described.
\end{finding}

\begin{finding}[Dataset accounting]\label{find:dataset}
The order-12 blocking file has 266 lines, not the previously reported 265 (the
file lacks a trailing newline), of which only 155 are distinct edge sets; 33
order-12 and 2 order-11 entries are disconnected, consisting of an
order-10/11 non-embeddable core plus isolated vertices. The blocking lists are
raw enumeration output, not a deduplicated minimal set. This is why
certificates carry induced-subgraph restriction, and why per-line bindings are
verified rather than assumed.
\end{finding}

\begin{finding}[The published final-candidate count is not
reproducible]\label{find:count}
The published sources disagree about the final order-22/23 candidates: the
paper's text reports one order-22 and 41 order-23 candidates surviving the
$\le 12$-subgraph filter, a figure caption implies only two candidates lack
minimal unembeddable subgraphs, and the published artifact's log for that step
has exactly two lines. We settled the question against the artifacts, with
SHA-256 hashes pinned: decoding the published exhaust files yields 88{,}282
distinct order-22 candidates and 5{,}160{,}001 distinct order-23 candidates;
replicating the pipeline's own filter semantics exactly (non-induced
monomorphism against its 17 minimal graphs) leaves 1 survivor at order 22 and
43 pairwise non-isomorphic survivors at order 23---not 41. Classifying the 43
against the located \uw{} minimal-unembeddable list (933 graphs) reconciles
the sources: exactly 41 contain an order-13 \uw{} minimal graph (seven
distinct patterns suffice), and the two that do not are precisely the pair in
the published log---so the caption's ``only two'' is correct with reference
class ``known minimal unembeddable graphs,'' while the ``41'' is not
reproducible from the artifact under the documented filter. None of this
threatens the bound; it does mean the verified chain must carry $\numSurvivors$
final candidates, not 42, and it is exactly the kind of discrepancy that
replayable proof objects exist to eliminate.
\end{finding}

Two further findings are internal to the checkers and illustrate the
discipline: substitution semantics were made unconditional (total-exponent
stripping) rather than relying on a well-formedness invariant the Lean checker
never verified, and monomial well-formedness was made normative at
deserialization in both checkers after the adversarial suite exposed a mutant
distinguishable only syntactically.

% =========================================================================
\section{Toward a Fully Verified Bound}\label{sec:fullbound}

The verified architecture has four layers, composing to ``no KS system on
$\le 23$ vectors,'' hence the bound $24$: \Lone{} (real geometry to
combinatorics: vector systems, orthogonality graphs, squarefreeness,
noncolorability, minimality WLOGs), \Ltwo{} (combinatorics to CNF:
encoding-correctness theorems for the generator's constraint set as shipped),
\Lthree{} (CNF exhaustiveness: verified replay of the enumeration's
unsatisfiability results together with a proved-sound canonicity predicate),
and \Lfour{} (candidate non-embeddability: this paper). \Lfour{} is closed for
the blocking database and, via the order-13 certificates, for $42$ of the
$\numSurvivors$ final candidates; the \Lone{} lemmas consumed so far are
proved.

Two design decisions dominate the remainder. For \Lthree{} canonicity,
trusting the solver's symmetry-breaking code is not acceptable, and the
published proof format resolves the choice favorably: its CAS-derived clauses
carry permutation witnesses for both the orderly-generation blocking and the
minimal-non-embeddable-subgraph blocking. Verifying a blocked candidate then
means applying the recorded permutation and checking a lexicographic
comparison, or checking a subgraph embedding against a certified
non-embeddable graph---both squarely within the verified toolkit of
\cref{sec:lean}. An alternative route, re-running the enumeration with VeriPB
dominance-based proof logging bridged into Lean \cite{GochtNordstrom2021,
PBLean2026}, is cleaner but costs compute; it remains the fallback if witness
regeneration proves brittle.

For \Lthree{} unsatisfiability, scale forbids kernel-level replay: the
published DRAT/LRAT material runs to 6.1\,TiB through order 22 and 40.3\,TiB
at order 23. The plan is a verified \emph{streaming} LRAT
checker---Lean-proved sound, compiled, constant-memory, in the tradition of
verified SAT proof checking \cite{CruzFilipe2017,Lammich2020}---composed with
the Lean-side statement, with only per-cube summary artifacts entering the
kernel.

The staging is deliberately tiered: first a fully self-contained re-derived
bound at small orders (a verified $\ge 22$, matching \cite{UijlenWesterbaan2016},
with verified canonicity and the complete lemma inventory), then replay
through order 23 for a fully verified $\ge 24$, contingent on the published
cube certificates or re-runs and a cluster budget. The composed theorem will
carry the two unresolved order-23 candidates as explicit hypotheses until
their direct certificates land; under the repository's invariants, an
unresolved graph is reported as unresolved, never absorbed.

% =========================================================================
\section{Related Work}\label{sec:related}

\paragraph{The minimum KS problem.}
The bound $\ge 18$ of Arends et al.\ \cite{ArendsOuaknineWampler2011}
introduced the graph-theoretic minimality arguments (squarefreeness, WLOG
constraints) the modern pipelines inherit; Uijlen and Westerbaan
\cite{UijlenWesterbaan2016} reached $\ge 22$ and published the minimal
non-embeddable graph lists we reuse at order 13; Li et al.\
\cite{LiBrightGanesh2024} and Kirchweger et al.\
\cite{KirchwegerPeitlSzeider2023} independently reached $\ge 24$ with
SAT-plus-computer-algebra pipelines. Our work audits and complements the
former's published pipeline and artifacts; nothing here diminishes those
results---the point is to make them independently checkable.

\paragraph{Verified SAT-based mathematics.}
Machine-checked replay of SAT-derived mathematical results is by now a
tradition, from the Boolean Pythagorean triples proof onward, resting on
verified DRAT and LRAT checkers \cite{CruzFilipe2017,Lammich2020} and,
recently, on pseudo-Boolean proof logging with dominance reasoning
\cite{GochtNordstrom2021} bridged into proof assistants \cite{PBLean2026},
whose encoding-correctness pattern our \Ltwo{} instantiates on a problem with
a fifty-year pedigree. The distinctive difficulty here is that SAT proofs
cover only half the argument; the other half lives in $\exists\Real$.

\paragraph{Certificates for real-algebraic reasoning.}
Sum-of-squares and Positivstellensatz certificates for real infeasibility are
classical in verified optimization and formal proof, and formally real fields
provide exactly the leverage our SOS splits use. Our contribution on this axis
is a certificate format tuned to orthogonality systems---WLOG anchoring,
schedule replay, injectivity obligations, and a positivity leaf over products
of pair-injectivity polynomials---small enough for a total checker with a
kernel-checked soundness proof and kernel-reducible verdicts. On the
satisfiable side of $\exists\Real$ search, recent work attacks realizability
with numerical-symbolic methods \cite{KrapivinPrzybockiHeule2026}; our
certificates target the unsatisfiable side with replayable objects, adjacent
but not overlapping.

\paragraph{KS formalizations.}
An existing Lean development proves the Kochen--Specker paradox itself via one
explicit vector gadget \cite{Gupta2025}---the existence direction. To our
knowledge, no prior machine-checked certificate exists for the geometric half
of any KS lower bound.

% =========================================================================
\section{Conclusion}\label{sec:conclusion}

The geometric half of the minimum Kochen--Specker bound no longer needs to be
taken on faith for the graphs the published proof consults at every step: all
$\numSourceLines$ source lines of the blocking database now carry exact
rational case-tree certificates, replayed by two independent checkers and
backed by closed Lean kernel theorems under a minimal axiom closure, at a cost
of about half a second of kernel time per graph. The exercise also
demonstrated, once more, why proof objects matter beyond replication: building
a checker one must prove sound surfaced a load-bearing injectivity condition,
hidden WLOG branches, an optimization misclassified as a lemma, and an
unreproducible headline count---none visible to a workflow that trusts its
solvers.

The path to a fully verified $\ge 24$ is scoped and, we believe, unblocked:
witness-based canonicity checking matches the published proof format,
streaming LRAT replay addresses the terabyte-scale SAT material, and $42$ of
the $\numSurvivors$ final candidates are already discharged by verified
subgraph containment. The remaining two candidates are tracked as explicit,
machine-readable obligations---which is, in the end, the entire method: never
a verdict without an object, never an object without a checker, never a
checker without a proof.

\paragraph{Artifact availability.}
Certificates, both checkers, the Lean development, the adversarial suite, and
the artifact-provenance records (including SHA-256 pins for
Finding~\ref{find:count}) are available at
\url{https://github.com/shayaansiddique06/kscert} under the Apache 2.0
license, with a pinned toolchain and a single-command build for each
component.

\paragraph{Acknowledgments.}
We thank the authors of \physcheck{} for publishing their pipeline and proof
artifacts, without which this audit-and-certify effort would not have been
possible. The development was carried out with assistance from Claude
(Anthropic).

% =========================================================================

\end{document}